\documentclass[preprint,showpacs,preprintnumbers,nofootinbib,amsmath,amssymb,aps,prc,longbibliography]{revtex4-1}
\usepackage{graphicx}% Include figure files
\usepackage{dcolumn}% Align table columns on decimal points
\usepackage{bm}% bold math
\begin{document}
%%%%%%%%%%%%%%%%%%%%%%%%%%%%%%%%%%%%%%%%%%%%%%%%%
\title{Giant dipole resonance in $^{201}$Tl at low temperature}
\author{N. Dinh Dang$^{1}$}
  \email{dang@riken.jp}
\author{N. Quang Hung$^{2}$}
 \altaffiliation[On leave of absence from Center for Theoretical and Computational Physics, ]
 {College of Education, Hue University, Vietnam}
 \email{hung.nguyen@ttu.edu.vn}

 \affiliation{1) Theoretical Nuclear Physics Laboratory, RIKEN Nishina Center
for Accelerator-Based Science,
2-1 Hirosawa, Wako City, 351-0198 Saitama, Japan\\
and Institute for Nuclear Science and Technique, Hanoi, Vietnam\\
2) School of Engineering, TanTao University, TanTao University Avenue, TanDuc Ecity, Duc Hoa, Long An Province, Vietnam and 
Institute for Computational Science and Technology, 6 Quarter, Linh Trung Ward, Thu Duc District, HoChiMinh City, Vietnam}

\date{\today}% It is always \today, today,
\begin{abstract}
The thermal pairing gap obtained by embedding the exact solutions of the pairing problem into the canonical ensemble is employed 
to calculate the width and strength function of the giant dipole resonance (GDR) within the phonon damping model.
The results of calculations describe reasonably well the data for the GDR width as well as the GDR linearized strength function, recently 
obtained for $^{201}$Tl in the temperature region between
0.8 and 1.2 MeV, which other approaches that neglect the effect of non-vanishing thermal pairing fail to describe.

%%%%%%%%%%%%%%%%%%%%%%%%%%%%%%%%%%%%%%%%%%%%%%%%%
\end{abstract}

%Use showkeys class option if keyword
\maketitle
%%%%%%%%%%%%%%%%%%%%%%%%%%%%%%%%%%%%%%%%%%%%%%%%%
\section{Introduction}
\label{Intro}
Since the discovery of the giant dipole resonance (GDR) as a collective thermal excitation 
in highly excited (hot) nuclei~\cite{Newton}, many experiments were carried out in the last three decades to 
extract the GDR width and its shape (linearized strength function) as functions of nuclear temperature $T$ and angular momentum $J$. 
A recent compilation of GDR built on excited states is given in Ref. \cite{Schiller}. 
At present, the well-established systematics accumulated by measuring the $\gamma$ decays of various hot compound nuclei formed in heavy-ion fusion reactions and inelastic scattering of light particles on heavy targets has shown that the GDR width increases with temperature $T$ within the temperature region 1 MeV $\leq T\leq$ 3 -- 4 MeV.
It has also shown that the GDR width's increase with angular momentum $J$ becomes noticeable only at $J \geq $ 27 -- 30 $\hbar$ in heavy nuclei, whereas
its location (peak energy) remains mostly unchanged as $T$ and $J$ vary. 

Experimental studies often refer to the thermal shape fluctuation model (TSFM)~\cite{TSFM} as one of 
theoretical descriptions of the width's increase in this temperature region. The TSFM takes the thermal average of 
the GDR photoabsorption cross section over the shape-dependent cross sections caused by 
all fluctuating quadrupole shapes, which are assumed to be coupled to the GDR vibration. 
The width's increase as a function of $T$ arises as a results of such thermal average.
The TSFM, however, fails to describe the temperature dependence of the GDR width beyond the temperature region 1.5 $< T\leq$ 3 MeV. At $T>$ 3 -- 4 MeV several experimental evidences have shown that the GDR width seems to saturate at high $T$~\cite{saturation}, whereas 
the TSFM predicts a continuously increasing width. In the low temperature region, at $T\leq$ 1 MeV, a measurement of $\gamma$ decays in coincidence with $^{17}$O particles scattered inelastically from $^{120}$Sn~\cite{Heckman} has obtained a GDR width in $^{120}$Sn of around 4 MeV at $T =$ 1 MeV, that is smaller than its value of 4.9 MeV at $T =$ 0. This result and the existing systematics for the GDR width in $^{120}$Sn up to $T\simeq$ 1.5 MeV are significantly lower than the prediction by the TSFM.

Meanwhile, the GDR width as a function of temperature $T$  is well described by the phonon damping model (PDM)~\cite{PDM1a,PDM1b} in the entire region 0$\leq T\leq$ 5 - 6 MeV, including the increase in the width at $T\leq$ 3 MeV as well as the width saturation at high $T$. Within the PDM, the damping of GDR at $T\neq$ 0 is caused by coupling of the GDR to noncollective particle-hole ($ph$) and particle-particle ($pp$) [hole-hole ($hh$)] configurations. The coupling to $ph$ configurations exists even at $T=$ 0, and  leads to the quantal width  $\Gamma_Q$, whereas the thermal width $\Gamma_T$ arises owing to coupling to $pp$ and $hh$ configurations, which appear only at $T\neq$ 0 because of the distortion of the Fermi surface at $T\neq$ 0. In the low temperature region $T\leq$ 1 MeV, it has been shown within the PDM that thermal pairing plays a crucial role in reducing the GDR width in $^{120}$Sn~\cite{PDM2}. As a matter of fact, in finite systems such as nuclei, the pairing gap does not collapse at the critical temperature $T_c \simeq$ 0.57$\Delta(T=0)$ [$\Delta(T=0)$ being the pairing gap at $T=$ 0] as in the case of the superfluid-normal phase transition in infinite systems, but decreases monotonically as $T$ increases. This decrease of pairing tends to restore the Fermi surface, which is diffused in the presence of pairing, back to the sharp step-function distribution. This competes with the thermal smoothing of the Fermi surface, which increases with $T$. As a result of such competition, a compensation takes place, which leaves the GDR width almost unchanged or even decreases slightly at $T\leq$ 1 MeV.   At $T\geq$ 2 MeV, pairing becomes significantly smaller than its value at $T=$ 0, so the thermal distortion of the Fermi surface becomes dominant and the width starts to increase. 

Very recently, $\alpha$ induced fusion reactions were used to measure the GDR width at low temperature~\cite{alpha1,alpha2}. These reactions can describe temperature more precisely whereas the associated angular momentum in the mass region $A=$ 115 -- 121 is rather small ($\leq$ 24$\hbar$). The data extracted from these latest experiments for the GDR width in $^{119}$Sb at 0.98 $\leq T\leq$ 1.23 MeV~\cite{alpha1} are similar to those obtained previously for $^{120}$Sn, including the data point at $T=$ 1 MeV mentioned above, in good agreement with the prediction by the PDM  for the GDR width in $^{120}$Sn when thermal pairing is included. However, these experiments also reported the data for the GDR width in $^{201}$Tl~\cite{alpha2}, which were extracted within the temperature interval 0.8 $\leq T<$ 1.2 MeV. These values are significantly smaller than the prediction by the TSFM for the GDR width in $^{208}$Pb even after including the shell effect. The authors of Ref. \cite{alpha2} also made a comparison with the prediction by the PDM using the results for the GDR width in $^{208}$Pb. However $^{208}$Pb is a doubly closed shell nucleus, that is the neutron and proton pairing gaps are both zero, and, naturally, no pairing was taken into account in the PDM prediction for the GDR width in this nucleus, whereas $^{201}$Tl is an open-shell nucleus for both neutrons (N=120) and protons (Z=81). Therefore, an adequate comparison should be made with the prediction within the PDM for the GDR width of the same $^{201}$Tl nucleus, including the effects owing to thermal pairing of neutrons as well as protons. The aim of the present paper is to make such prediction.

As has been mentioned above, the conventional finite-temperature (FT) BCS theory, which exhibits a collapse of the pairing gap at a critical temperature $T_c$, should be modified  to include thermal fluctuations when it is applied to finite nuclei. Among such modifications are the modified BCS (MBCS)~\cite{MBCS}, the finite-temperature BCS1 (FTBCS1) as well as the Lipkin-Nogami projected FTBCS1 (the so-called FTLN1)~\cite{FTBCS1}. However, these approaches fail for a close-to-magic nucleus such as $^{201}$Tl (Z=81). Therefore, in the present paper, we employ the exact treatment of thermal pairing within the canonical ensemble (CE), which has been elaborated in Ref. \cite{CE}. Within this approach, the thermal pairing gaps are calculated from the exact pairing energy, which is obtained by averaging the exact eigenvalues of the pairing problem in the CE at temperature $T$. This approach also allows us to calculate the exact single-particle occupation numbers, chemical potentials, as well as the exact quasiparticle energies. 
By using the latter, one can determine the quantities that approximate the coefficients $u_k$ and $v_k$ of the Bogolyubov's transformation from particles to quasiparticles as well as the corresponding quasiparticle occupation numbers and use them as ingredients to calculate the GDR width as a function of $T$ in $^{201}$Tl within the quasiparticle representation of the PDM. Because the recent data for the GDR width in $^{201}$Tl at low $T$ were obtained at low angular momentum (below 25$\hbar$), the effect of angular momentum on the GDR width is negligible. Therefore, for simplicity, angular momentum is not included in the present calculations (See Ref.~\cite{PDMJ} for the recent extension of PDM to finite angular momentum.).

The paper is organized as follows. The formalism is derived in Sec. \ref{formalism}. The results of numerical calculations are analyzed and compared with the experimental data in Sec. \ref{results}. The paper is summarized in the last section, where conclusions are drawn.
%%%%%%%%%%%%%%%%%%%%%%%%%%%%%%%%%%%%%%%%%%%%%%%%%
\section{Formalism}
\label{formalism}
%%%%%%%%%%%%%%%%%%%%%%%%%%%%%%%%%%%%%%%%%%%%%%%%%
\subsection{GDR width within the PDM including thermal pairing}
%%%%%%%%%%%%%%%%%%%%%%%%%%%%%%%%%%%%%%%%%%%%%%%%%
Because the PDM has been discussed in great details in a series of papers~\cite{PDM1a,PDM1b,PDM2,PDMJ,PDM,viscosity,PDMmore}, we summarize below only the main results necessary for the numerical calculations in the present paper.

The PDM considers a model Hamiltonian, which consists of three terms. The first term describes the independent single-particle 
(quasiparticle) field with single-particle (quasiparticle) energies $\epsilon_k$ ($E_k$), the second term stands for the phonon field with phonon energies 
$\omega_q$, whereas last term treats the 
coupling between these two fields [See Eq. (1) in Ref. \cite{PDM1a} for example.]. 
This coupling causes the damping of the phonon vibrations, for example, the GDR phonon ($q=GDR$).
As a result, the GDR acquires a width. The expression of the GDR full width at half maximum (FHWM) $\Gamma$ is obtained within the PDM as 
\begin{equation}
\Gamma(T) = 2\gamma_q(\omega=E_{GDR})~,
\label{FWHM}
\end{equation}
where $\gamma_q(\omega=E_{GDR})$ is the phonon damping at the GDR peak energy $E_{GDR}$. 
In the presence of superfluid pairing, the phonon damping $\gamma_q(\omega)$ has the explicit form as~\cite{PDM2}
\begin{equation}
\gamma_q(\omega)=\pi\bigg\{{\cal F}_1^2\sum_{ph}[u_{ph}^{(+)}]^{2}(1-n_{p}-n_{h})
\delta(\omega-E_{p}-E_{h})+{\cal F}_2^2\sum_{ss'}[v_{ss'}^{(-)}]^{2}(n_{s'}-n_{s})
\delta(\omega-E_{s}+E_{s'})\bigg\}~.
\label{gamma}
\end{equation}
In this expression, $(ss')$ stands for $(pp')$ and $(hh')$ with $p$ and $h$ denoting 
the particle ($p$) levels, that is those situated above the Fermi level at $T=$ 0 and $\Delta=$ 0, 
and hole ($h$) levels,  that is those situated below it. 
${\cal F}_1$ and ${\cal F}_2$ are the model parameters for $ph$ and $pp$ ($hh$) couplings, respectively.
Functions $u_{ph}^{(+)}$ and $v_{ss'}^{(-)}$ are 
combinations of the Bogolyubov's coefficients
$u_{k}$, $v_{k}$, namely $u_{ph}^{(+)}=u_{p}v_{h}+v_{p}u_{h}$, and
$v_{ss'}^{(-)}=u_{s}u_{s'}-v_{s}v_{s'}$. The first sum at the right-hand side of Eq. (\ref{gamma}) with the factors
$(1-n_{p}-n_{h})$ represents quantal damping caused by coupling of the GDR phonon to noncollective $ph$ configurations, 
whereas the second sum with the factors
$(n_{s'}-n_{s})$ stands for thermal damping, which arises from coupling of 
the GDR phonon to $pp$ and $hh$ configurations. Because $\omega\geq$ 0, the second sum over $(ss')$ is finite (and positive) only at $E_s>E_{s'}$. 
The quasiparticle occupation number $n_{k}$ ($k= p, h$) is given in the form of
a Fermi-Dirac distribution
\begin{equation}
n_{k}=[\exp(E_{k}/T)+1]^{-1},
\label{nk}
\end{equation}
smoothed with a Breit-Wigner-like kernel, 
whose width is equal to the quasiparticle damping with the quasiparticle energy
\begin{equation}
E_{k}=\sqrt{[\epsilon_{k}-\lambda(T)]^{2}+\Delta(T)^{2}}~,
\label{Ek}
\end{equation}
where $\lambda(T)$ and $\Delta(T)$ denote the temperature-dependent chemical potential and pairing gap, respectively 
 [See Eq. (2) of Ref. \cite{PDM2}.].  For the GDR in medium and heavy nuclei, the quasiparticle damping is usually small, therefore 
the Breit-Wigner-like kernel can be safely replaced with the $\delta$-function. As a result, the quasiparticle occupation number  
 $n_{k}$ has the form of the Fermi-Dirac
 distribution (\ref{nk}). This also means that thermal damping of the GDR appears only at $T\neq$ 0 because all $n_k$ vanish at $T=$ 0, whereas quantal damping exists already at $T=$ 0, when the factors $(1-n_p - n_h) =$ 1, as well as at $T\neq$ 0. In closed-shell nuclei, the pairing gap $\Delta(T)$ is zero, so we have $(1-n_p-n_h)=f_h-f_p$, $(n_{p'}-n_p)=f_{p'}-f_p$, and $n_{h'}-n_{h}=f_{h}-f_{h'}$, where $f_k$ are the single-particle occupation numbers. Correspondingly, the sums and differences of quasiparticle energies become $E_p+E_h = \epsilon_p-\epsilon_h$,
$E_{p}-E_{p'}=\epsilon_p-\epsilon_{p'}$, and $E_{h}-E_{h'}=\epsilon_{h'}-\epsilon_{h}$. As a result, Eq. (\ref{gamma}) reduces to the expression for the phonon damping for the case without pairing~\cite{PDM1a,PDM1b}, that is
\begin{equation}
\gamma_q(\omega) = \pi\sum_{kk'}[F_{kk'}^{(q)}]^{2}(f_k-f_{k'})\delta(\omega-\epsilon_{k'}+\epsilon_k)~,
\label{gamma_nopair}
\end{equation}
with $F_{ph}^{(q)}={\cal F}_1^0$ and $F_{pp'}^{(q)} = F_{hh'}^{(q)}={\cal F}_2^0$, where ${\cal F}_{1,2}^{0}$ denote the corresponding values of parameters ${\cal F}_{1,2}$ in the zero-pairing case ($\Delta = 0$). As has been demonstrated in Ref. \cite{PDM2} (See the discussion below Fig. 3 therein), the factors $[u_{ph}^{(+)}]^2(1-n_p-n_h)$ cause a slight decrease in the quantal width as $T$ increases, whereas the factors $[v_{ss'}^{(-)}]^2(n_{s'}-n_s)$ are responsible for the strong increase in the thermal width with $T$ at low and moderate $T$, and its saturation at high $T$. The combined effect makes the total width increase with $T$ at low and moderate $T$ and saturate at high $T$. At very low $T$ ($T\leq$ 1 MeV) pairing effect leaves the total width almost unchanged with $T$ or, in some cases, makes it even smaller than the value at $T=$ 0, as has been discussed in the Introduction.
 
The escape width $\Gamma^{\uparrow}$, which arises because of coupling to the continuum and is related to the direct decay by particle emission, is usually small (few hundred keV), and does not seem to the change with $T$  in medium and heavy nuclei. In the numerical calculations within the PDM, its effect is taken into account via the smoothing parameter $\varepsilon$, which replaces the $\delta$-functions in Eqs. (\ref{gamma}) and (\ref{gamma_nopair}) with $\delta(x) \rightarrow \varepsilon/[\pi(x^{2}+\varepsilon^2)]$, where $x=\omega-E_{p}-E_{h}$ and $\omega-E_{s}+E_{s'}$ in Eq. (\ref{gamma}), and $x = \omega-\epsilon_{k'}+\epsilon_k$ in Eq. (\ref{gamma_nopair}). A value of $\varepsilon=$ 0.5 MeV is adopted in the present calculations. The results do not change significantly with $\varepsilon$ varying between 0.5 and 1 MeV. The PDM does not take into account the evaporation width $\Gamma_{ev}$ of the compound nuclear states, which comes from the quantum mechanical uncertainty principle~\cite{evaporation} because its effect on the GDR width is expected to be significant only at high $T$ ($\gg$ 3.3 MeV) and $J$ ($\gg$ 30$\hbar$).

The GDR strength function $S(\omega)$ is calculated as
\begin{equation}
S(\omega) =\frac{1}{\pi}\frac{\gamma_q(\omega)}{[\omega-E_{GDR}]^2+\gamma_q^2(\omega)}~.
\label{S}
\end{equation}
The GDR energy $E_{GDR}$ is found as the solution of the equation [See Eqs. (3) and (4) of Ref. ~\cite{PDM2}.]:
\begin{equation}
\omega-\omega_q-\bigg\{{\cal F}_1^2\sum_{ph}\frac{[u_{ph}^{(+)}]^{2}(1-n_{p}-n_{h})}
{\omega-E_{p}-E_{h}}+{\cal F}_2^2\sum_{ss'}\frac{[v_{ss'}^{(-)}]^{2}(n_{s'}-n_{s})}
{\omega-E_{s}+E_{s'}}\bigg\}= 0~,
\label{EGDR}
\end{equation}
where the expression within the figure brackets is the real part of the polarization operator $P_q(E)$ at $E=\omega\pm i\varepsilon$, which causes 
the energy shift of the phonon energy $\omega_q$ under the effect of quasiparticle-phonon coupling, whereas the imaginary part of 
$P_q(\omega\pm i\varepsilon)$ is the phonon damping in Eq. (\ref{gamma}). 
%%%%%%%%%%%%%%%%%%%%%%%%%%%%%%%%%%%%%%%%%%%%%%%
\subsection{Exact treatment of thermal pairing within the canonical ensemble}
%%%%%%%%%%%%%%%%%%%%%%%%%%%%%%%%%%%%%%%%%%%%%%%%
\label{CE}
As for the pairing problem, the present work considers the exact treatment of thermal pairing within the canonical ensemble (CE) with the pairing Hamiltonian given as
\begin{equation}
H_P=\sum_k\epsilon_k(a_{+k}^{\dagger}a_{+k}+a_{-k}^{\dagger}a_{-k}) - G\sum_{kk'}{a_k^{\dagger}
a_{-k}^{\dagger}a_{-k'}a_{k'}}~, 
\label{Hpair}
\end{equation}
where $a_{\pm k}^{\dagger}(a_{\pm k})$ are the creation (annihilation) operators 
of a particle (neutron or proton) having the single-particle energy $\epsilon_k$, 
angular momentum $j_k$ with projections $m_k>0$, denoted with $+k$, and with projection $-m_k<0$, denoted with $-k$.
The exact eigenvalues ${\cal E}_s$ of the eigenstates $s$ with degeneracies $d_s$, 
obtained by diagonalizing this Hamiltonian, are used to construct the partition function within the CE at temperature $T$~\cite{CE}:
\begin{equation}
    {Z}(\beta)=\sum_{s}d_{s}e^{-\beta{\cal E}_{s}}~,\hspace{5mm} \beta=1/T~. 
    \label{ZCE}
    \end{equation}
By using this CE partition function, the total exact CE energy ${\cal E}$, entropy $S$, and occupation
number $f_{k}^{CE}$ on the $k$th level are obtained in terms of
the ensemble averages of the exact eigenvalues ${\cal E}_s$ and state-dependent occupation
numbers $f_k^{(s)}$, respectively, as
\begin{equation}
    {\cal E}=-\frac{\partial\ln Z(\beta)}{\partial\beta},\hspace{5mm}   S= \beta{\cal E}
     +\ln Z(\beta)~, \hspace{5mm} 
   f_{k}^{CE}=\frac{1}{Z(\beta)}\sum_{s}f_{k}^{(s)}d_s e^{-\beta{\cal E}_s}~.      
     \label{Ef}
\end{equation}
Knowing the total energy ${\cal E}(n)$ and entropy $S(n)$ of a system with $n$ particles (n = N neutrons or Z protons), we can calculate its Helmholtz free energy 
$F(n)={\cal E}(n) - TS(n)$ and the exact CE chemical potential $\lambda^{CE}(T)$ as~\cite{chemical}
\begin{equation}
\lambda^{CE}(T) = \frac{1}{4}[F(n+2) - F(n-2)]~.
\label{lambda}
\end{equation}
This approach does not produce a pairing gap, which is a mean field concept. Instead, an exact CE pairing gap is introduced to 
mimic the mean-field pairing gap as [See Eq. (18) of Ref. \cite{CE} and the discussion therein.], namely:
\begin{equation}
    \Delta^{CE}(T)=\sqrt{-G{\cal E}_{\rm pair}}~,\hspace{5mm}
    {\cal E}_{\rm pair}={\cal E}-{\cal E}^{(0)}~, \hspace{5mm} {\cal
    E}^{(0)}\equiv
    2\sum_{k}(\epsilon_{k}- \frac{G}{2}f_{k}^{CE})f_{k}^{CE}~.
    \label{gap}
    \end{equation}    
It is worth mentioning that, although the CE approach to thermal pairing was employed by several authors~\cite{Delft,Ross,Frau,Suma}, 
the gap (\ref{gap}) is very similar to that defined in Eq. (52) of Ref. \cite{Delft}, but 
it is different from the canonical gap defined in Refs. \cite{Ross,Frau} because 
the term $\langle{\cal E}\rangle_{\rm C}^{(0)}$ is taken at
$G=$ 0 in the latter, whereas no thermal pairing gap was calculated in Ref. \cite{Suma}.

Given the chemical potential $\lambda^{CE}(T)$ in Eq. (\ref{lambda}) and the pairing gap $\Delta(T)$ in Eq. (\ref{gap}), the quasiparticle energies $E_k$ can be calculated by using Eq. (\ref{Ek}). In the same way the exact CE pairing gap is introduced to mimic the mean-field pairing gap,  the quantities
that mimic the Bogolyubov's coefficients $u_k$ and $v_k$ can be obtained separately for neutrons and protons by using the standard expressions 
\begin{equation}
u_k = \sqrt{\frac{1}{2}\bigg[1+\frac{\epsilon_k-\lambda(T)}{E_k}\bigg]}~,\hspace{5mm}      
v_k = \sqrt{\frac{1}{2}\bigg[1-\frac{\epsilon_k-\lambda(T)}{E_k}\bigg]}~,
\label{uv}
\end{equation}
which, strictly speaking, are valid only within the BCS-based theories. A justification for such approximation is that, in the cases where both the exact CE gaps and the FTBCS1 as well as the FTLN1 are possible (including the corrections owing to coupling to the self-consistent quasiparticle random-phase approximation), these pairing gaps are rather close to each other, especially at $T\leq$ 2 MeV (See Fig. 1 of Ref. \cite{CE}.).
%%%%%%%%%%%%%%%
\section{Analysis of numerical results}
%%%%%%%%%%%%%%%
\label{results}
The 323 neutron and proton doubly folded single-particle energy levels for $^{201}$Tl employed in the present calculations are obtained within the axially deformed Woods-Saxon potentials including the spin-orbit and Coulomb interactions~\cite{WS}. The neutron spectrum spans a space of 157 doubly folded levels (degenerated into 41 spherical orbitals), starting from the bottom at -38.36 MeV up to 12.57 MeV. The proton spectrum has 166 doubly folded  levels (degenerated into 44 spherical orbitals), starting from the bottom at around -34 MeV, up to around 22.3 MeV.  They cover the energy intervals similar to those used in the calculations of GDR width in $^{208}$Pb~\cite{PDM1a,PDM1b}.

In the construction of the CE partition function (\ref{ZCE}) one needs to include all the eigenvalues of the ground state as well as excited states. Meanwhile,  the {\scriptsize FORTRAN IMSL} subroutine for matrix diagonalization at the RIKEN Integrated Cluster of Clusters computing system implies that the numbers of levels $\Omega$ and particles $n$ should satisfy the condition $\Omega!/[(n/2)!(\Omega-n/2)!] < 10^4$, which means it is impossible to carry out the exact diagonalization of the pairing Hamiltonian (\ref{Hpair}) with the entire single-particle spectra~\cite{CE}.
Therefore, knowing that pairing has a significant effect around the Fermi surface, we calculate the exact CE thermal gap (\ref{gap}) only for 14  doubly folded neutron (proton) levels situated around the Fermi one, $\lambda(T=0)$, with 7 levels below and 7 levels above it.  The selected levels belong to the group of 18 neutron (proton) levels within five $(2j+1)$-folded spherical orbitals 
$1i_{13/2}, 3p_{3/2}, 2f_{5/2}, 3p_{1/2}, 2g_{9/2}$ for the neutrons, and $1h_{11/2}, 2d_{3/2}, 3s_{1/2}, 1h_{9/2}$, $2f_{7/2}$ for protons, as listed in Table~\ref{table1}. The pairing parameters $G_N=$ 0.182 MeV for neutrons  and $G_{Z}$ = 0.37 MeV for protons are chosen to reproduce the empirical values of the neutron and proton gaps $\Delta_{N,Z}$, which are both equal to 1 MeV for $^{201}$Tl according to Fig. 2-5 of Ref. \cite{Bohr}. This value agrees well with the three-point and five-point gaps calculated in Ref. \cite{Bender} and shown in Fig. 1 therein. The remaining 4 neutron levels (two on $1i_{13/2}$ and two on $2g_{9/2}$ orbitals) and 4 proton levels (two on $1h_{11/2}$ and two on $2f_{7/2}$ orbitals) of this group are assumed to have the same thermal neutron and proton pairing gaps, respectively. For the levels beyond this group, pairing is assumed to be negligible so that $1-n_h = f_h$, $n_p = f_p$, $u_{p(h)}=1~(0)$, and $v_{p(h)}= 0 ~(1)$. 
%%%%%%%%%%%%%%%%%%%%%%%%%%%%%% Table 1
\begin{table}
\begin{center}
    \caption
    {Single-particle spherical orbitals obtained within Woods-Saxon potentials and used in the calculations of neutron and proton exact CE  pairing gaps for $^{201}$Tl. \label{table1}}
    \vspace{2mm}
\begin{tabular}{|c|cc|cc|}
\hline
& \multicolumn{2}{c|}{$N$} & \multicolumn{2}{c|}{$Z$}\\
\hline k &
\multicolumn{1}{c}{~~~$nl_j$~~~}&\multicolumn{1}{c|}{~~~$\epsilon_k$~(MeV)~~~}&\multicolumn{1}{c}{~~~$nl_j$~~~}&\multicolumn{1}{c|}{~~~$\epsilon_k$~(MeV)~~~}
\\ \hline 
1&$2g_{9/2}$&-2.226&$2f_{7/2}$&-3.766\\
2&$3p_{1/2}$&-5.620&$1h_{9/2}$&-3.957\\
3&$2f_{5/2}$&-6.189&$3s_{1/2}$&-8.038\\
4&$3p_{3/2}$&-6.586&$2d_{3/2}$&-8.342\\
5&$1i_{13/2}$&-6.903&$1h_{11/2}$&-10.031\\

\hline
\end{tabular}
\end{center}
\end{table}
%%%%%%%%%%%%%%%%%%%%%%%%%%

The PDM assumes that the matrix elements of GDR coupling to $ph$ configurations, causing the quantal width, are all equal to ${\cal F}_1$,  whereas those of coupling to $pp$ ($hh$) configurations, causing the thermal width at $T\neq$ 0, are all equal to ${\cal F}_2$ [See Sec. II B of Ref. \cite{PDM2} for the detail discussion on the justification of this assumption.]. The third parameter, $\omega_q$, in the case of GDR ($q=1$), is chosen to be close to $E_{GDR}(T=0)$. Because the mechanism of the spreading width $\Gamma^{\downarrow}$ at $T=$ 0 is known, which is owing to 
coupling to more complicate configurations $2p2h$ configurations, the PDM has no ambition to calculate it microscopically but is interested only in its temperature dependence incorporated in the quantal width. Therefore the parameter ${\cal F}_1$ is selected to reproduce the GDR experimental width at $T=$ 0, which is essentially the sum of the spreading width, $\Gamma^{\downarrow}$, and the escape width, $\Gamma^{\uparrow}$. The parameter ${\cal F}_2$ is usually adjusted so that the GDR energy $E_{GDR}(T)$, found as the solution of Eq. (\ref{EGDR}), does not change appreciably with $T$.  Because the GDR energy in $^{201}$Tl does not depend on $T$, namely $E_{GDR}(T)=13.8$ MeV~\cite{alpha2}, for simplicity, we adopt in the present paper $E_{GDR} =$ 13.8 MeV, and select ${\cal F}_2$ so that the calculated width at $T\simeq$ 2 MeV matches the corresponding experimental value for the GDR width in $^{208}$Pb~\cite{Baumann}.

%%%%%%%%%%%%%%%%%%%%%%%%%% Fig. 1
    \begin{figure}
       \includegraphics[width=13.5cm]{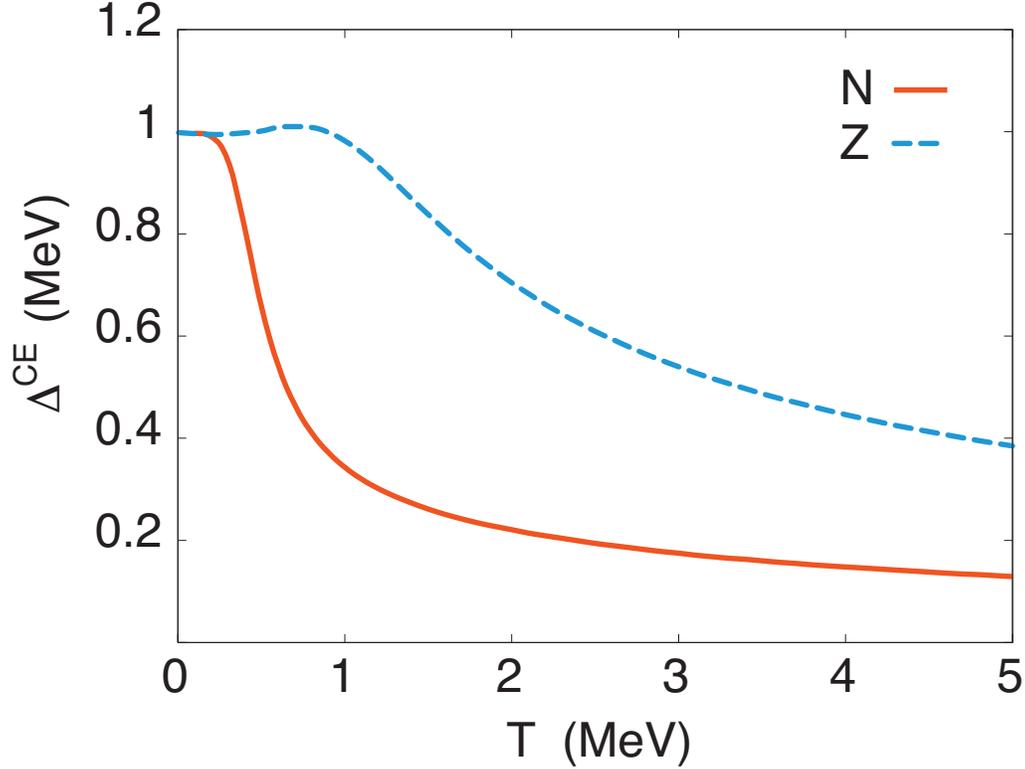}
        \caption{(Color online)  Exact CE gaps for neutrons [(red) solid line] and protons [(blue) dashed line] for $^{201}$Tl.
        \label{gapTl201}}
    \end{figure}
%%%%%%%%%%%%%%%%%%%%%%%%%%%%%%%%%%%%%%%%%%%%%%%%%
Shown in Fig. \ref{gapTl201} are the exact CE  neutron and proton pairing gaps for $^{201}$Tl as functions of $T$. In different with the BCS gaps, the exact CE gaps
do not collapse at $T_c\simeq$ 0.57 MeV for $\Delta(0)=$ 1 MeV. Instead, they both decrease as $T$ increases. In particular, the proton gap remains almost unchanged or even slightly increases in the region $0\leq T\leq $ 1 MeV, and it is significantly larger than the neutron gap at the same value of $T$ already starting from $T\simeq$ 0.5 MeV. At $T=$ 1 MeV the proton gap is still equal to around 1 MeV whereas the neutron gap drops to 0.34 MeV. At $T=$ 5 MeV the proton gap remains as large as around 0.4 MeV, whereas the neutron gap is depleted to 0.13 MeV.

%%%%%%%%%%%%%%%%%%%%%%%%%% Fig. 2
    \begin{figure}
       \includegraphics[width=11.0cm]{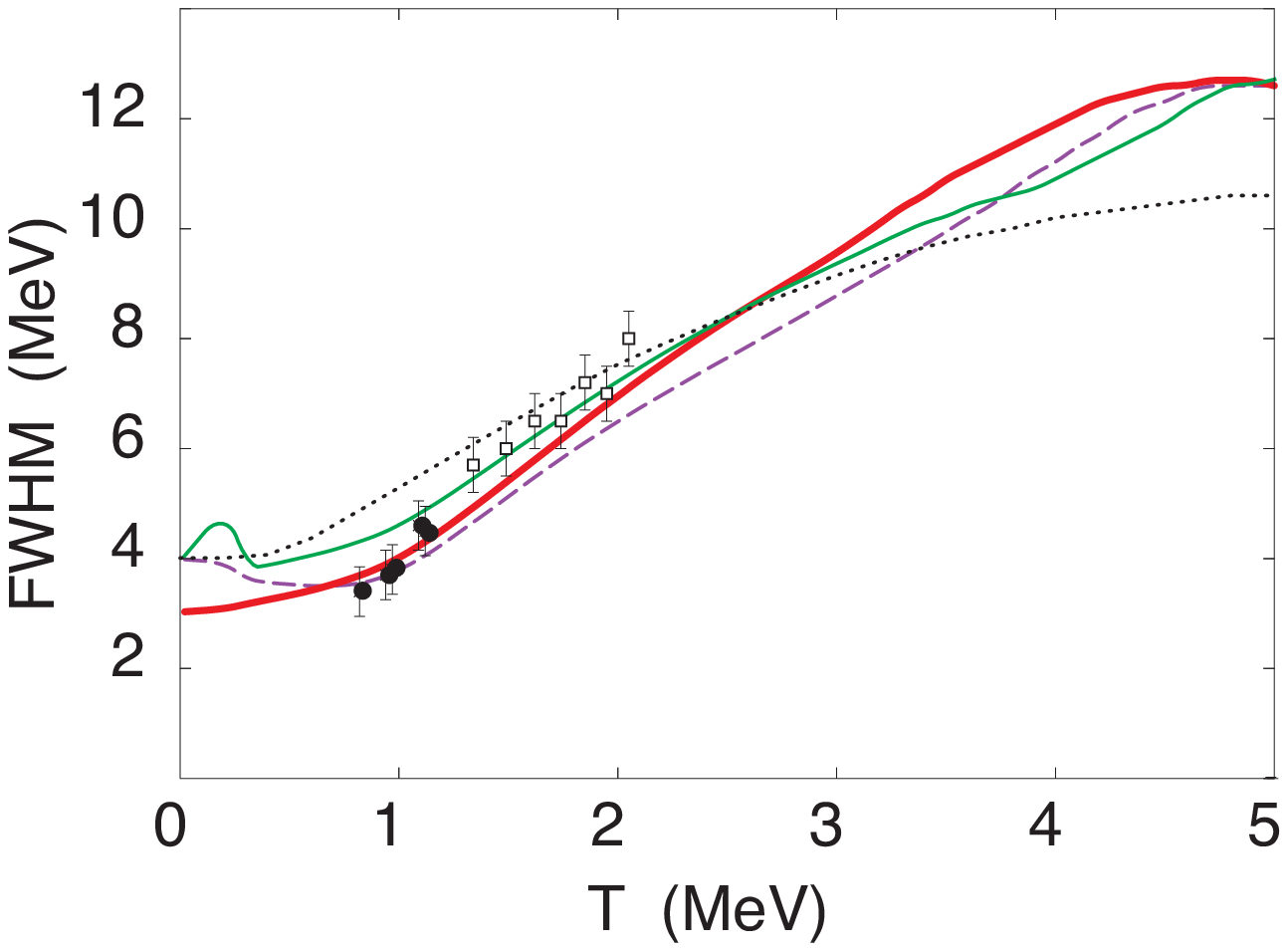}
        \caption{(Color online)  GDR width as a function of $T$ versus experimental data. The (red) thick solid line is the GDR width for $^{201}$Tl obtained with the effect of exact CE pairing. The (green) thin solid line denotes the same width but obtained without pairing.  The (purple) dashed line stands for the same quantity, but obtained by using the exact single-particle occupation numbers $f_k$ and effective $\epsilon_k-\lambda(T)$ discussed in the text. The dotted line represents the GDR width for $^{208}$Pb obtained previously in Refs. \cite{PDM1a,PDM1b}. The full circles and open boxes are  experimental GDR widths for $^{201}$Tl~\cite{alpha2} and $^{208}$Pb~\cite{Baumann}, respectively.
        \label{widthTl201}}
    \end{figure}
\begin{figure}
       \includegraphics[width=11cm]{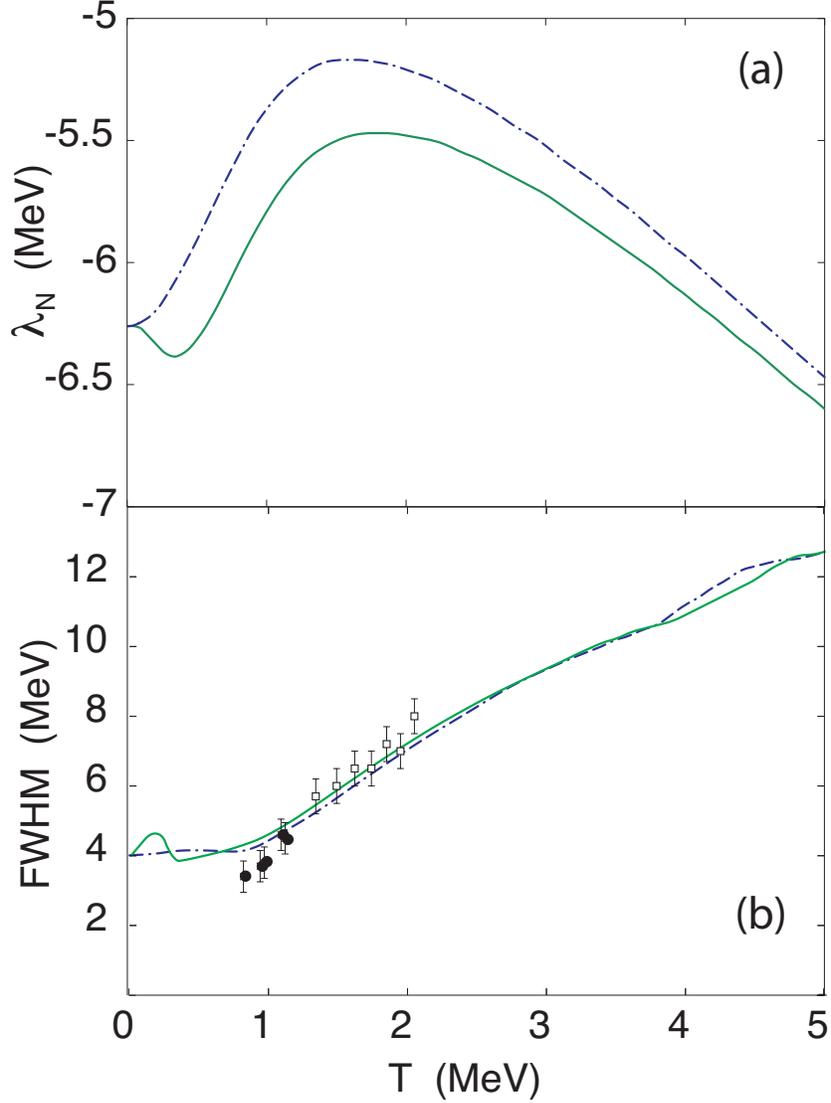}
        \caption{(Color online)  Effect of the single-particle orbitals at low $T$. In (a) the (green) solid line denotes the neutron chemical potentials as a function of $T$ obtained for $^{201}$Tl  before modifying the energies of the orbitals $i_{13/2}$ and $p_{3/2}$, whereas the (blue) dash-dotted line stands for the same quantity obtained after modifying the energies of these orbitals as specified in the text. In (b) the (green) solid and (blue) dash-dotted lines show the GDR widths for $^{201}$Tl obtained within the no-pairing approximation before and after this modification, respectively. 
                \label{widthtest}}
    \end{figure}

%%%%%%%%%%%%%%%%%%%%%%%%%%%%%%%%%%%%%%%%%%%%%%%%%
The GDR FWHM $\Gamma(T)$ in $^{201}$Tl obtained within the PDM in three different approximations are plotted as functions of $T$
against the experimental data in Fig. \ref{widthTl201}. In the first approximation, the zero-pairing one, the width $\Gamma(T)$ is calculated by using Eq. (\ref{FWHM}) and the phonon damping given in Eq. (\ref{gamma_nopair}). The single-particle occupation numbers $f_k$ are approximated with the Fermi-Dirac distribution 
\begin{equation}
f_k = 1/\{\exp[\beta(\epsilon_k-\lambda)]+1\}~,
\label{fFD}
\end{equation}
where the chemical neutron and proton potentials change with $T$ to conserve the particle numbers, according to the equation $n = 2\sum_k f_k$ with $n = N, Z$. The calculations adopted the following values of the parameters for the $ph$ and $pp$ ($hh$) couplings: ${\cal F}_1=$ 4.0$\times$10$^{-2}$ MeV and ${\cal F}_2\simeq$ 13.27$\times$10$^{-2}$ MeV. The GDR width predicted by this approximation (thin solid line) clearly overestimates three lowest experimental data points at 
$T=$ 0.82, 0.84 and 0.97 MeV. The overall temperature dependence of this width is similar to that obtained previously for the doubly closed-shell   $^{208}$Pb (dotted line) except at $T>$ 3 MeV, where it becomes larger than the GDR width in $^{208}$Pb, to saturate at $T>$ 4 MeV at a value of 12.7 MeV compared to that of around 10.5 MeV for $^{208}$Pb.  The small bump at $T<$ 0.3 MeV occurs because of the neutron single-particle energies of the orbitals $i_{13/2}$ and $p_{3/2}$, which are equal to 6.903 MeV and 6.587 MeV, respectively, and located just below the Fermi level at $T=$ 0. They cause a local minimum in the temperature dependence of the neutron chemical potential at $T=$ 0.3 MeV as shown by the solid line in Fig. \ref{widthtest} (a). A test by changing the energies of these levels to 7.903 and 7.587 MeV, respectively, removes this local minimum and, consequently, flattens the GDR width at $T<$ 1 MeV, as shown by the dash-dotted lines in Figs. \ref{widthtest} (a) and \ref{widthtest} (b), respectively.
%%%%%%%%%%%%%%%%%%%%%%%%%% Fig. 3
%%%%%%%%%%%%%%%%%%%%%%%%%%%%%%%%%%%%%%%%%%%%%%%%%

In the second approximation, we consider an effective way of taking thermal pairing into account, namely, in Eq. (\ref{gamma_nopair}), the single-particle occupation numbers $f_k$ for the 18 selected levels around the Fermi surface, which were discussed previously, are replaced with their exact CE values $f_k^{CE}$ calculated by using Eq. (\ref{Ef}). The effective values of $(\epsilon_k -\lambda)$  for these 18 levels are found by inverting the Fermi-Dirac distribution (\ref{fFD}) to obtain 
$(\epsilon_k -\lambda) = T [\ln(1-f_k^{CE})-\ln f_k^{CE}]$. The calculations used ${\cal F}_1\simeq$ 2.45$\times$10$^{-2}$ MeV and ${\cal F}_2\simeq$ 14.83$\times$10$^{-2}$ MeV. Starting at the same value equal to around 4 MeV at $T=$ 0, 
the width predicted by this approximation (dashed line) decreases as $T$ increases up to $T=$ 0.7 MeV to perfectly match the three lowest 
data points at $T=$ 0.82, 0.84 and 0.97 MeV. At $T>$ 0.7 MeV this width increases with $T$ but remains smaller than that obtained without pairing up to
$T\simeq$ 3.6 MeV when they cross and reach the same value at $T=$ 5 MeV. 

Finally, thermal pairing is fully taken into account within the quasiparticle representation of the PDM by using Eqs. (\ref{FWHM}) and (\ref{gamma}), which include the exact CE thermal pairing gap for 18 levels around the Fermi surface for neutrons and protons, separately, as has been discussed in Sec. \ref{CE}. The GDR width predicted by this approach (thick solid line) nicely describes all the data points for $^{201}$Tl. Its values at $T =$ 0.8, 0.9, 1.0, 1.1 and 1.2 MeV are found equal to 3.58, 3.75, 3.96, 4.22 and 4.5 MeV, respectively, to be compared with the corresponding experimental values of 
3.4$\pm$ 0.45, 3.7$\pm$ 0.45, 3.8$\pm$ 0.45, 4.6$\pm$ 0.45 and 4.5$\pm$ 0.45 MeV at $T=$ 0.82, 0.94, 0.97, 1.09 and 1.12 MeV, respectively. It also matches fairly well the experimental width for $^{208}$Pb at 1.3$<T\leq$ 2 MeV, although, up to $T\simeq$ 3 MeV, it remains smaller than the width predicted by the PDM for $^{208}$Pb (dotted line). However, to obtain such description, the value for ${\cal F}_1$ has to be smaller than that used in the case without pairing so that $\Gamma(0)\simeq$ 3 MeV instead of 4 MeV for the GDR width in $^{208}$Pb at $T=0$, and, consequently, the value for  ${\cal F}_2$ has to be larger to reproduce the same data point at $T=$ 2 MeV. Eventually, this brings $\Gamma(0)$ closer to the recent parametrization of the GDR width for quasipherical nuclei~\cite{Junghans}, which implies that  the GDR FWHM for $^{201}$Tl at $T=$ 0 amounts to around 3.33 MeV, rather than 4 MeV as that of $^{208}$Pb. The slope of the width's increase as a function of $T$ also becomes slightly steeper, improving the agreement between theory and experiment within the entire region of $T$. A value for ${\cal F}_1$ that reproduces $\Gamma(0)=$ 4 MeV leads to an overestimation of the width in the entire temperature region of $T\neq$ 0. The values for the coupling parameters used in this approximation are 
${\cal F}_1\simeq$ 2.81$\times$10$^{-2}$ MeV and ${\cal F}_2\simeq$ 16.73$\times$10$^{-2}$ MeV.  It is important to point out that, although one could fit the data points for $^{201}$Tl within the zero-pairing approximation by reducing the value of parameter ${\cal F}_1$ from 4.0$\times$10$^{-2}$ MeV to 3.54$\times$10$^{-2}$ MeV, it is not possible by readjusting only parameters ${\cal F}_{1,2}$ to achieve the overall agreement with both sets of data for the GDR widths in $^{201}$Tl and $^{208}$Pb as does the last approach that fully includes exact CE pairing gaps within the quasiparticle representation of the PDM. At $T>$ 2.5 MeV the GDR width obtained by using such reduced parameter ${\cal F}_1$ in the zero-pairing approximation becomes even smaller than that predicted by the second approximation (of taking effectively thermal pairing into account). This indicates that the effect owing to thermal pairing on the GDR width has its microscopic origin, which cannot be accounted for by simply adjusting model's parameters.  

%%%%%%%%%%%%%%%%%%%%%%%%%% Fig. 4
    \begin{figure}
       \includegraphics[width=10cm]{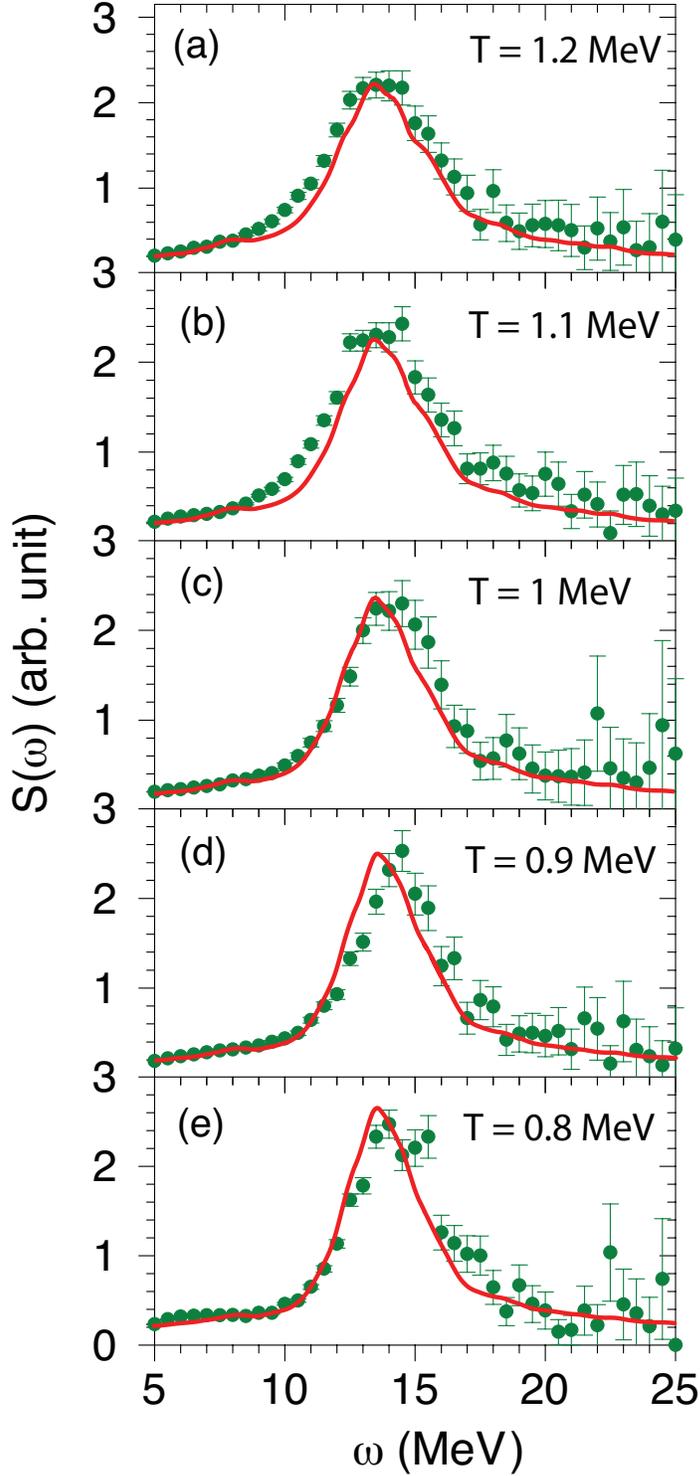}
        \caption{(Color online)  GDR strength functions for $^{201}$Tl obtained within the quasiparticle representation of the PDM (red solid lines) at $T =$ 0.8 -- 1.2 MeV as shown in the panels in comparison with the experimental linearized strength functions (green data points)~\cite{alpha2} at $T=$ 0.82, 0.94, 0.97, 1.09, and 1.12 MeV from the bottom panel, (e), to the top one, (a), respectively.
        \label{stre}}
    \end{figure}
%%%%%%%%%%%%%%%%%%%%%%%%%%%%%%%%%%%%%%%%%%%%%%%%%
The GDR strength functions obtained at $T=$ 0.8 -- 1.2 MeV by using Eq. (\ref{S}) within the quasiparticle representation of the PDM that includes exact CE pairing are displayed in Fig. \ref{stre} in comparison with the corresponding experimental data adapted from Fig. 1(a) of Ref. \cite{alpha2}. The theoretical strength functions have been renormalized so that the value at $\omega=$ 5 MeV and the maximum at $T =$ 1.2 MeV match the corresponding experimental values. This figure and Fig. \ref{widthTl201} show that the PDM describes reasonably well not only the temperature dependence of the GDR width but also that of the GDR linearized shape.

%%%%%%%%%%%%%%%%%%
\section{Conclusions}
%%%%%%%%%%%%%%%%%%%%
In the present paper, we calculated the width and strength function of the GDR in $^{201}$Tl at finite temperature within the framework of the quasiparticle representation of the PDM. Thermal pairing is taken into account by using the exact treatment of pairing within the canonical ensemble.
This treatment allows us to calculate the exact equivalences to the pairing gaps for protons and neutrons in a nucleus neighboring a proton closed-shell one. Because of thermal fluctuations owing to the finiteness of the system, which are inherent in the CE, the exact CE thermal pairing gaps do not collapse at the critical temperature $T_c$ of the superfluid-normal phase transition as in the case of infinite systems, but decrease monotonically as $T$ increases, and remain finite up to $T$ as high as 5 MeV. The theoretical predictions within the PDM are compared with the data, which were recently obtained for the GDR width and strength functions in $^{201}$Tl at 0.8$<T<$ 1.2 MeV at low angular momentum below 25$\hbar$.

The good agreement between the PDM predictions including thermal pairing and the experimental data is a clear demonstration of the manifestation of the effect owing to thermal pairing, which plays a vital role in reducing the GDR width at low $T$ in open-shell nuclei. Under the influence of thermal pairing, the GDR width in $^{201}$Tl becomes as low as around 3.7 MeV at $T=$ 0.8 MeV, and the width $\Gamma(0)$ of the GDR built on the ground state ($T=$ 0) can be as small as 3 MeV, that is smaller than the GDR width in $^{208}$Pb (4 MeV) at $T=$ 0. The results obtained in the present work as well as the previous predictions for the GDR width in $^{120}$Sn, where the important role of neutron thermal pairing has been shown to reduce the GDR width at $T\leq$ 1 MeV~\cite{PDM2}, confirm that, in order to have an adequate description of GDR damping at low $T$, a microscopic model needs to take into account thermal pairing at least up to $T\sim$ 1.5 MeV.   
%%%%%%%%%%%%%%%%%%%%%%%%%%%%%%%%%%%%%%%%%%%%
\acknowledgments
The numerical calculations were carried out using the {\scriptsize FORTRAN IMSL}
Library by Visual Numerics on the RIKEN Integrated Cluster of Clusters (RICC) system. 
The authors are grateful to S.R. Banerjee for useful discussion and help with the experimental data of GDR linearized strength functions for $^{201}$Tl.
NQH acknowledges the support by the National Foundation for Science and Technology Development(NAFOSTED) of Vietnam through Grant No. 103.04-2010.02. 
%%%%%%%%%%%%%%%%%%%%%%%%%%%%%%%%%%%%%%%%%%%%

%%%%%%%%%%%%%%%%%%%%%%%%%%%%%%%%%%%%%%%%%%%%%%%%%
\end{document}